\documentclass[superscriptaddress,floatfix,twocolumn,showpacs,preprintnumbers,amsmath,amssymb,aps,prl]{revtex4-1}

\usepackage{sidecap} 

\usepackage{graphicx}
\usepackage{subfigure}
\usepackage{dcolumn}
\usepackage{bm}

\begin{document}

\title{Generation of Coherent Extreme-Ultraviolet Radiation Carrying Orbital Angular Momentum}

\author{Primo\v{z} Rebernik Ribi\v{c}}
\author{David Gauthier}
\affiliation{Sincrotrone Trieste, Trieste 34149, Italy}
\author{Giovanni De Ninno}
\affiliation{Sincrotrone Trieste, Trieste 34149, Italy}
\affiliation{Laboratory of Quantum Optics, University of Nova Gorica, Nova Gorica 5001, Slovenia}

\date{\today}

\begin{abstract}
We propose an effective scheme for the generation of intense coherent extreme ultraviolet light beams carrying orbital angular momentum (OAM). The light is produced by a high-gain harmonic-generation free-electron laser (FEL), seeded using a laser pulse with a transverse staircase-like phase pattern. During amplification, diffraction and mode selection drive the radiation profile towards a dominant OAM mode at saturation. With a seed laser at 260 nm, gigawatt power levels are obtained at wavelengths approaching those of soft x-rays. Compared to other proposed schemes to generate OAM with FELs, our approach is robust, easier to implement, and can be integrated into already existing FEL facilities without extensive modifications of the machine layout.
\end{abstract}

\maketitle

\section{}
Modern generation free-electron lasers (FELs) delivering high-brightness optical beams in the extreme ultraviolet (XUV) \cite{allaria:2012} and x-ray regions \cite{ackermann:2007,emma:2010} have become indispensable tools for probing structural and chemical properties of matter at femtosecond temporal and nanometer spatial resolutions \cite{ribic:2012}. At present, transverse radiation profiles from FELs working at saturation are limited to a fundamental Gaussian-like mode with no azimuthal phase variation. This is true for FELs based on self-amplified spontaneous emission (SASE), where the amplification starts from electron shot-noise \cite{kim:1986,chin:1992,xie1:2000,xie2:2000,xie:2001,huang:2007,saldin:2000,saldin:2001,saldin1:2008,saldin2:2008}, as well as for seeded FELs, such as those based on high-gain harmonic-generation (HGHG), where the amplification process is triggered by a coherent input signal \cite{yu:2000,huang:2000,allaria:2012}.

Generation of high-order radiation modes, however, is a subject of strong interest, not only from the fundamental point of view but also in practical applications. In particular, helically phased light beams or optical vortices with a field dependence of $\exp{(il\phi)}$, where $\phi$ is the azimuthal coordinate and $l$ an integer referred to as the topological charge, are currently among intensively studied topics in optics. These light beams, which carry orbital angular momentum (OAM) \cite{allen:1992} that can be transferred to atoms, molecules, and nanostructures \cite{picon1:2010,jauregui:1992,babiker:1992,alexandrescu:1992,babiker:2002,toyoda:2013}, have already been utilized at visible and infrared wavelengths in a wide variety of applications, ranging from micromanipulation \cite{he:1995}, detection of spinning objects \cite{lavery:2013}, microscopy \cite{jesacher:2005}, and optical data transmission \cite{wang:2012,cai:2012,bozinovic:2013}. Perhaps the most promising applications of vortex beams at short wavelengths are in x-ray magnetic circular dichroism, where different OAM states allow the separation of quadrupolar and dipolar transitions \cite{veenendaal:2007}, photoionization experiments, where the  dipolar selection rules are violated giving rise to new phenomena beyond the standard effect \cite{picon2:2010}, and in resonant inelastic x-ray scattering, where vortex-beam-mediated coupling to vibrational degrees of freedom could provide important information on a wide range of molecular materials \cite{rury:2013}.

In the case of visible light, OAM is commonly generated by sending the beam through a suitable optical element (e.g., a spiral phase plate). This technique has been used in the past to produce XUV or x-ray beams that carry OAM \cite{terhalle:2011,peele:2002}. However, for high-brightness short-wavelength FEL radiation, the damage threshold of optical elements placed into the beam path, and the difficulties in the fabrication of high quality optical surfaces, impose strong limitations on the use of this method. Therefore, approaches using \textit{in situ} optical vortex generation are preferred.

In the case of undulator radiation, pioneering theoretical work by Sasaki and McNulty showed that x-ray OAM beams are produced as higher harmonics in a helical undulator \cite{sasaki:2008}, a principle recently demonstrated in an experiment \cite{bahrdt:2013}. For FELs, Hemsing and coworkers proposed two clever approaches to generate vortex beams at short wavelengths. The first one exploits the interaction of an electron beam (e-beam) with a seed laser in a helical undulator \cite{hemsing:2011}, while the second one is based on the echo-enabled harmonic generation (EEHG) scheme \cite{stupakov:2009}, where two seed lasers and two magnetic chicanes are used to produce harmonic microbunching of an e-beam with a corkscrew distribution \cite{hemsing:2012}. A proof-of-principle experiment has recently been performed to demonstrate the first scheme using a single undulator section, generating optical vortices at 800 nm \cite{hemsing:2013} . In this approach, however, OAM beams are produced at the fundamental frequency of the seed. Reaching short wavelengths would therefore require a coherent XUV or x-ray input signal, which is not trivial to obtain. On the other hand, the technique based on EEHG uses a relatively complex setup, which has yet to be thoroughly tested in experiments. 

In this letter, we show that a relatively simple setup, based on the original HGHG scheme \cite{yu:1991}, can be exploited in order to generate optical vortices at high harmonics of the seed laser, with wavelengths approaching those of soft x-rays. The method should, in principle, be straightforward to implement at existing seeded FEL user facilities, without the need for major machine upgrades.

\begin{figure}
\includegraphics[width=0.45\textwidth]{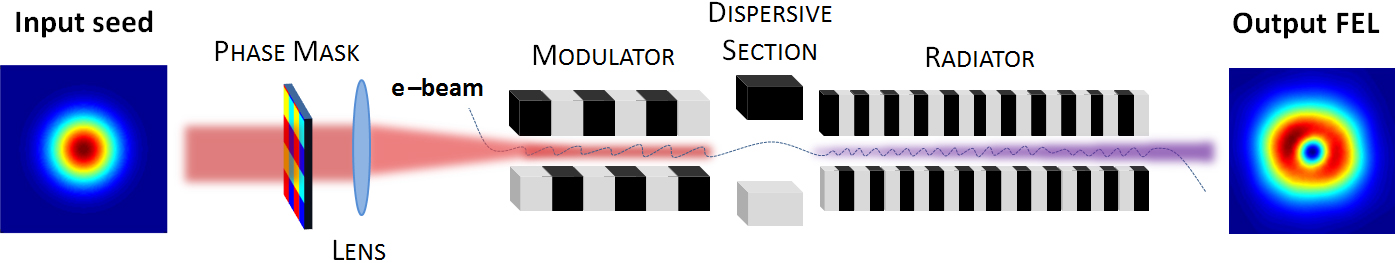} 
  \caption{The scheme to generate XUV OAM beams using a HGHG free-electron laser.}
  \label{scheme}
\end{figure}

The scheme is shown in Fig. \ref{scheme}. The main difference with respect to the standard HGHG setup \cite{yu:1991} is the use of an optical phase mask in order to create a transverse phase modulation in the seed laser profile. Naively, the simplest way to produce an XUV/x-ray optical vortex with this setup would be to seed the FEL directly with an OAM beam, by using a spiral phase plate as the phase mask \cite{beijersbergen:1994}. However, this approach fails at short wavelengths. To see why, let us look at the HGHG process in more detail. During the interaction of the seed laser with the e-beam in the first undulator, called the modulator, the seed properties, including the transverse helical phase dependence, are imprinted onto the e-beam as an energy modulation. Therefore, the longitudinally microbunched beam coming out of the dispersive section, which follows the modulator, also carries the OAM phase signature. The spatial microbunching in the longitudinal direction contains significant harmonic components. However, the topological charge $l_n$ of higher harmonics is multiplied with the harmonic number $n$ \cite{hemsing:2012}; i.e., $l_n=ln$, where $l$ is the topological charge of the seed. This results in a high-order OAM mode at the entrance of the second undulator, called the radiator, which is tuned to $\lambda=\lambda_s/n$, where $\lambda_s$ is the seed laser wavelength. Due to a lower coupling with the e-beam and stronger diffraction, this high-order OAM mode is not amplified in the radiator \cite{hemsing:2012,hemsing:2008}, leading to a dominant fundamental (non-OAM) mode at saturation.

The idea behind our approach is the following: instead of a helical transverse phase profile, a four quadrant staircase-like phase structure is imprinted onto an axially symmetric e-beam in the modulator. The resulting transverse distribution of electrons in phase can be represented by the following matrix: 
\begin{equation}
B_m=\begin{bmatrix}
\frac{1}{2}\pi & 0 \\
\pi & \frac{3}{2}\pi
\end{bmatrix} \mbox{,}
\label{b_matrix_modulator}
\end{equation}
meaning simply that the electrons with the azimuthal coordinate between $-\pi/4$ and $\pi/4$ have a relative phase of $0$, the electrons with $\pi/4 \leq \phi < 3\pi/4$ have a relative phase of $\pi/2$ and so on. Following frequency up-conversion at the radiator entrance, the phase distribution is multiplied by $n$, giving:
\begin{equation}
B_r= n \begin{bmatrix}\frac{1}{2}\pi & 0 \\ \pi & \frac{3}{2}\pi \end{bmatrix} \mod 2\pi \mbox{,}
\label{b_matrix_radiator}
\end{equation}
which for odd $n=2k+1$, where $k$ is an integer, becomes:
\begin{equation}
B_r= \begin{cases} \begin{bmatrix} \frac{1}{2}\pi & 0 \\ \pi & \frac{3}{2}\pi \end{bmatrix} = B_m, & \mbox{for even } k \\
\begin{bmatrix} \frac{3}{2}\pi & 0 \\ \pi & \frac{1}{2}\pi \end{bmatrix}, & \mbox{for odd } k \mbox{.} \end{cases}
\label{b_matrix_radiator2} 
\end{equation}
For the case of odd $k$, the elements in the main diagonal are interchanged, meaning that the transverse microbunching phase in the e-beam increases in the clockwise instead of anticlockwise direction.

The above equations show that the transverse microbunching structure is preserved for odd harmonics even after frequency up-conversion. The odd harmonics therefore carry the same staircase-like transverse phase pattern, which determines the spatial properties of the radiation at the radiator entrance. Because this initial bunching distribution contains a strong helical component, the radiation profile evolves into a dominant $l=1$ OAM mode at saturation. With an initial Gaussian transverse seed profile with $\lambda_s=260$ nm and e-beam parameters corresponding to modern seeded FELs, optical beams carrying orbital angular momentum at XUV wavelengths can be generated.  

\begin{figure}[h!]
\includegraphics[width=0.45\textwidth]{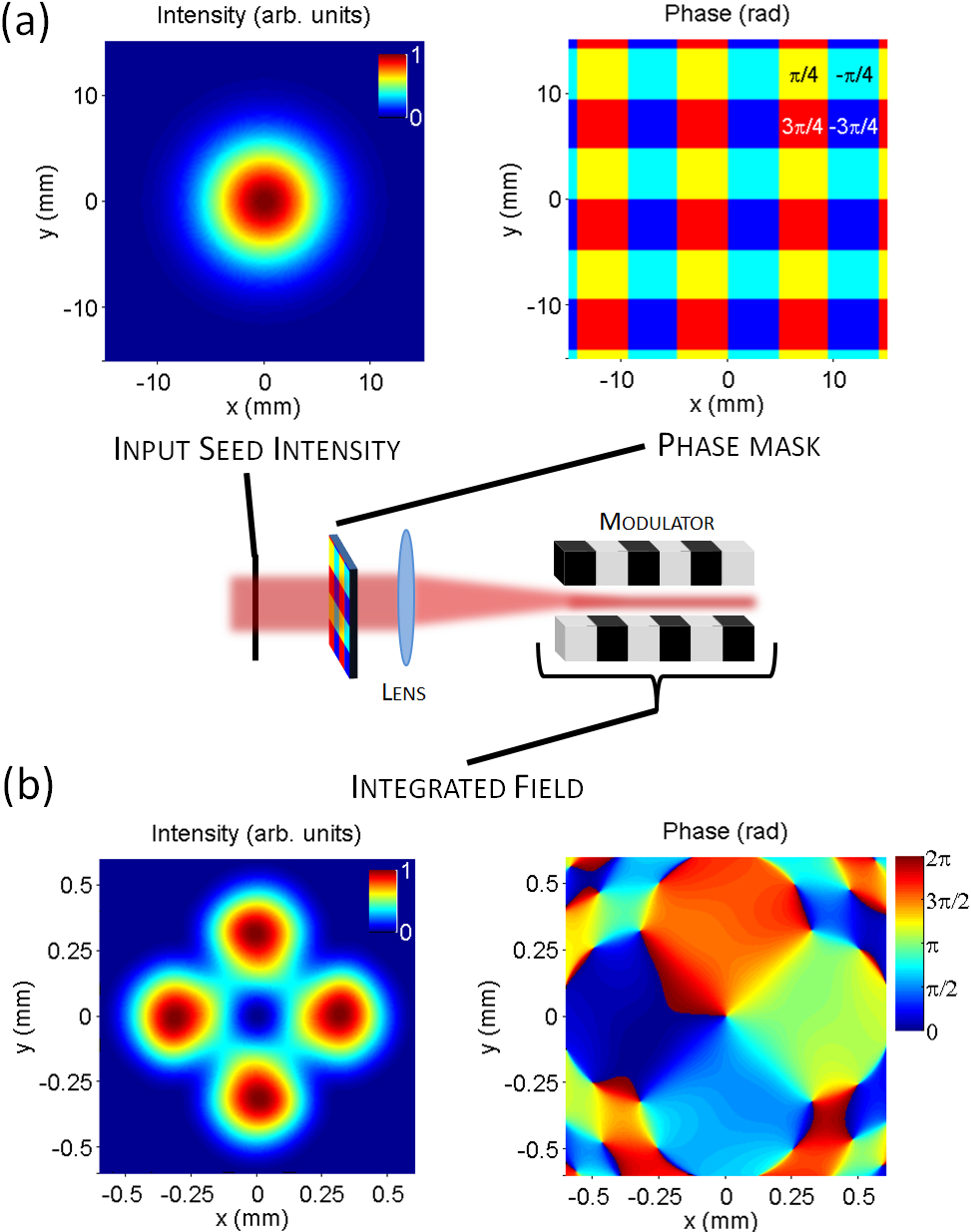} 
  \caption{a) The optical phase mask (right) used to modulate the transverse intensity profile of an initially Gaussian seed laser (left). b) Transverse intensity (left) and phase (right) distributions of the integrated seed laser field along the modulator.}
  \label{phase_mask}
\end{figure}

Fig. \ref{phase_mask} a) provides details on the phase mask inserted into the beam path of the seed laser with an initial Gaussian transverse intensity distribution. The lens, placed just after the phase mask, performs a Fourier transform of the transversely modulated electric field distribution at the focal plane, located in the middle of the modulator. This gives a transverse phase profile of the seed laser similar to the one represented by the matrix $B_m$. Due to diffraction, the staircase-like phase structure is maintained only near the focal plane of the seed. However, by choosing a relatively long focal distance (10 m), the integration of the seed laser electric field along the modulator, Fig. \ref{phase_mask} b), which is (in first approximation) proportional to the e-beam energy modulation, gives a phase profile which closely resembles that given by $B_m$. Despite the long focal distance, with seed laser powers around 1 GW or less, the bunching amplitude after the dispersive section is still sufficiently high to trigger the amplification of the $l=1$ OAM mode in the radiator.

\begin{figure}
\includegraphics[width=0.45\textwidth]{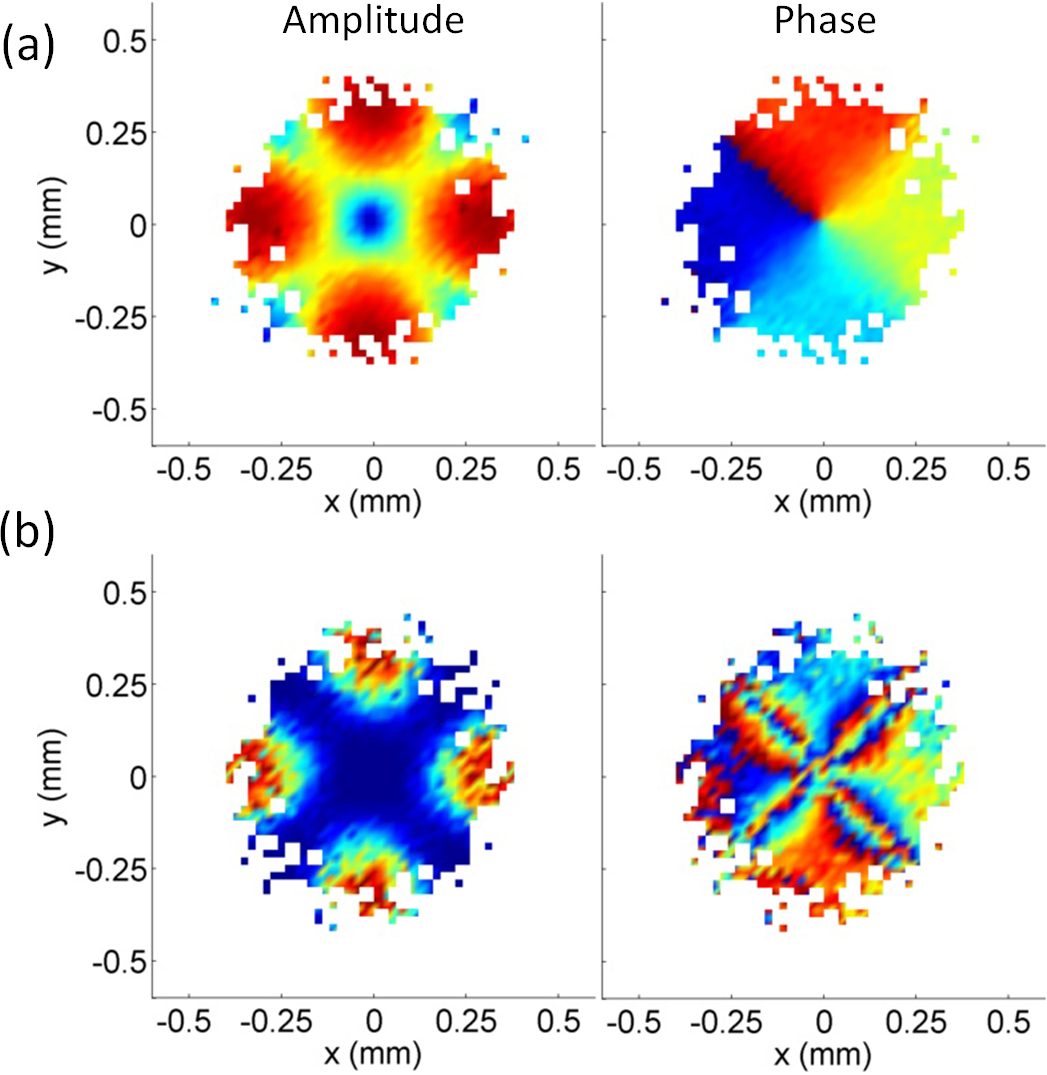} 
  \caption{Transverse microbunching construction in a) the modulator and dispersive section, and b) after frequency up-conversion, at the beginning of the radiator.}
  \label{bunching}
\end{figure}

The construction of microbunching in the modulator and FEL amplification in the radiator was studied in details using GENESIS \cite{reiche:1999}. A 1.3 GeV e-beam (0.1\% energy spread) with a transverse rms size of 150 $\mu$m, normalized emittance $\epsilon=5.0$ $\mu$m, and peak current of 1.5 kA was injected into a 1-m-long modulator tuned to the fundamental of the seed laser with $\lambda_s=260$ nm and power of 1 GW. Fig. \ref{bunching} a) shows the transverse dependence of the amplitude and phase (argument) of the local microbunching factor defined as $b(\vec{r})=<\exp{[i\theta_i(\vec{r})]}>$, where $\theta_i$ is the ponderomotive phase of the $i$-th particle and the brackets denote the ensemble average over all the particles at a certain transverse position $\vec{r}$ in the e-beam. The bunching distribution was evaluated after the beam passed through the dispersive section with $R_{56}= 50$ $\mu$m. The figure demonstrates that the transverse properties of the seed laser are efficiently transferred onto the e-beam. The microbunching distribution is maintained after frequency up-conversion when the e-beam enters the radiator, as shown in Fig.\ref{bunching} b). Due to the fact that the transverse microbunching phase profile is not perfectly flat inside the four quadrants, the phase structure also contains high frequency components. However, since the microbunching amplitude is relatively low in these areas, the generation of the OAM mode is not affected, as demonstrated in the following. 

\begin{figure}
\includegraphics[width=0.45\textwidth]{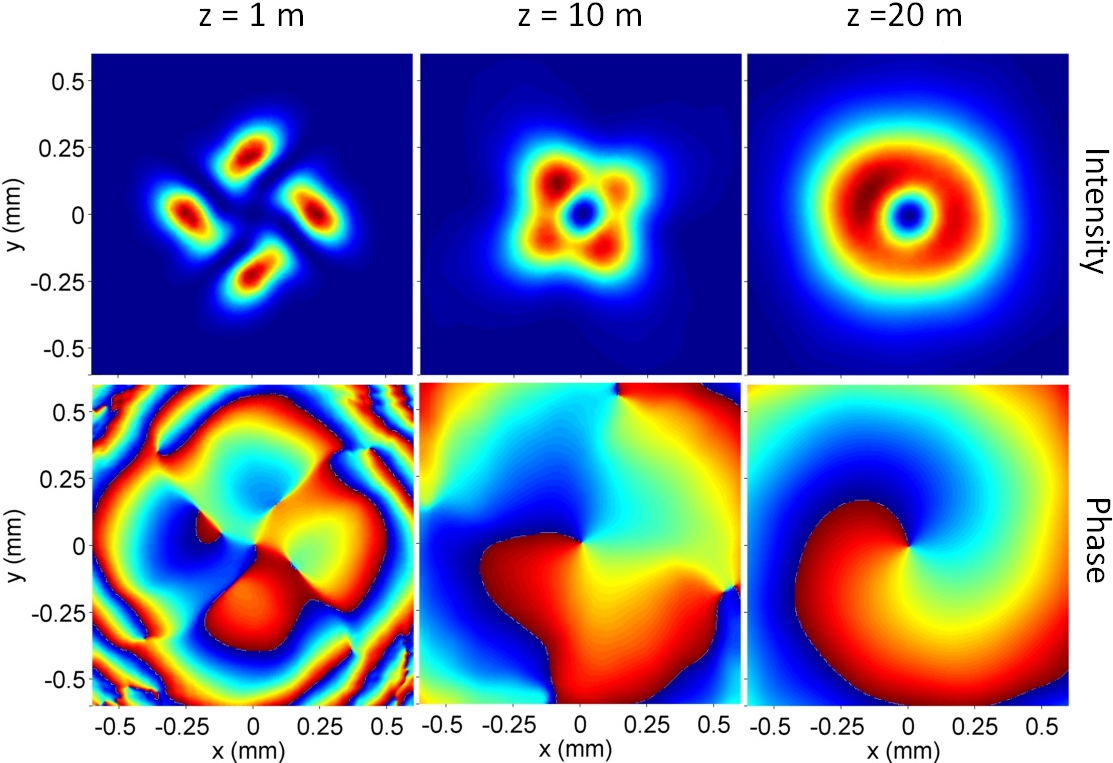} 
  \caption{Evolution of the transverse radiation profile in the radiator.}
  \label{radiation_profile}
\end{figure}

Fig. \ref{radiation_profile} shows the evolution of the radiation profile in a 20 m long radiator, tuned to the 7th harmonic of the seed laser, i.e. $\lambda=37.1$ nm. The initial field ($z=1$ m) has a phase and intensity structure determined by the microbunching distribution after frequency up-conversion (\textit{cf.} Fig. \ref{bunching} b)), clearly showing the four-quadrant staircase-like phase dependence. Since the initial radiation profile is not a guided FEL radiation mode with a self-similar intensity distribution, it will evolve due to diffraction and amplification in the radiator. Initially, at the radiator entrance, the bunching construction can be considered as rigid \cite{yu:1991}, therefore the evolution of the radiation profile is mainly governed by diffraction and linear amplification of the field. This is exemplified in the middle panel of Fig. \ref{radiation_profile} where the radiation profile has evolved from a four-quadrant/four-sources-like distribution to a profile highly resembling a ring-like intensity pattern with a helical phase dependence. As the e-beam enters the second half of the radiator, the FEL starts operating in the high-gain regime. Here, the transverse mode selection process \cite{kim:1986,chin:1992,xie1:2000,xie2:2000,xie:2001,huang:2007,saldin:2000,saldin:2001,saldin1:2008,saldin2:2008} favors radiation profiles of lower modes, such as the fundamental Gaussian-like and the $l=1$ OAM modes. Generally, in the high-gain regime and for a sufficiently long radiator, only the mode with the highest growth rate (Gaussian-like non OAM mode) will remain at saturation. However, using the seeding scheme described above, diffraction in the first part of the radiator gives rise to a strong $l=1$ mode at the beginning of the exponential growth regime. Therefore, the $l=1$ OAM mode reaches saturation well before the fundamental mode. As other higher modes are filtered out due to lower growth rates, the amplification process drives the radiation profile towards a dominant guided OAM mode with unit topological charge at saturation, as shown in the right panel of Fig. \ref{radiation_profile}.

\begin{figure}
\includegraphics[width=0.45\textwidth]{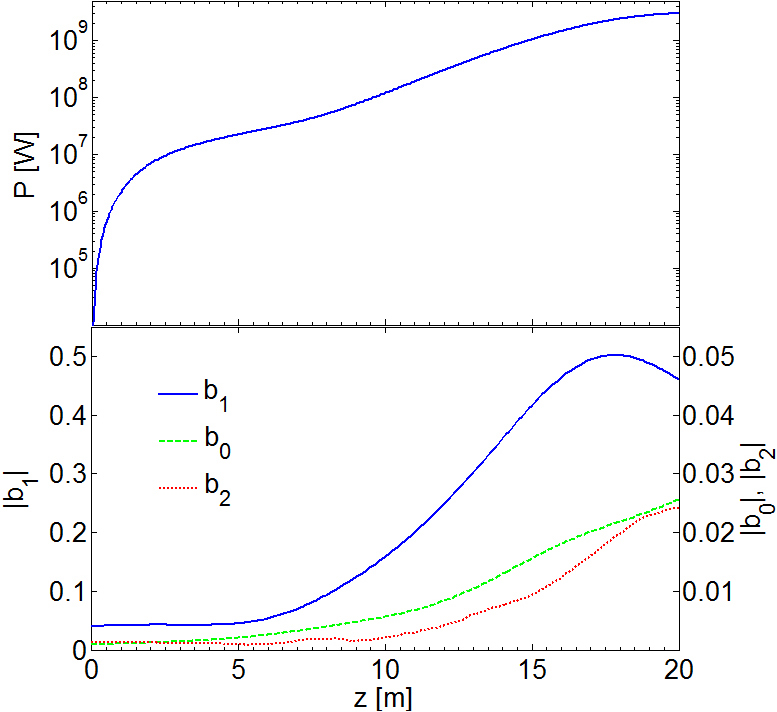} 
  \caption{Evolution of the FEL power (top) and the bunching factors (bottom) along the radiator.}
  \label{power}
\end{figure}

The above conclusions are supported by Fig. \ref{power} where we plot the power, modulus of the longitudinal bunching factor for the fundamental mode $b_0=<\exp{(i\theta_i)}>$, and the helical bunching factors for the two higher (OAM) modes with $l=1,2$, defined as $b_l=<\exp{(i\theta_i-il\phi)}>$ \cite{hemsing:2011,hemsing:2012}. Here the brackets denote the ensemble average over all the particles in the e-beam. The figures show the evolution of the radiation from the initial quadratic regime, where the bunching factors remain almost constant, into the high-gain regime where the modes are amplified. Throughout the amplification process, the bunching at the fundamental and $l=2$ modes remains below $\approx5$\% of the bunching factor $b_1$ corresponding to the $l=1$ mode. This shows that with our seeding scheme a strong $l=1$ component is excited at the radiator entrance. The intensity as well as $b_1$ attain maximum values at the radiator exit as the $l=1$ OAM mode reaches saturation at a gigawatt power level.

Figs. \ref{radiation_profile} and \ref{power} demonstrate the power and robustness of our approach. By simply using an optical phase mask in order to inject an e-beam with a strong helical bunching component into the radiator, an OAM mode with gigawatt power levels is obtained at saturation. Even though the seed laser does not imprint a pure $l=1$ OAM profile onto the e-beam, this mode emerges as the dominant one automatically at saturation due to diffraction and mode selection in the radiator.

Our method can be readily extended to higher harmonics. For shorter wavelengths, the radiator length at which a pure $l=1$ OAM mode is obtained scales with the Rayleigh range, which is proportional to $1/\lambda$, assuming a constant beam size of the seed laser. Nevertheless, by adjusting the modulator length, seed waist and power, optical vortices at wavelengths approaching those of soft x-rays can be generated efficiently with e-beam parameters of modern seeded FELs.

The above results were obtained for the case of a planar radiator. We have also performed simulations using a helical radiator where, due to a shorter gain length, we could relax the condition on the electron peak current required to reach saturation (1 kA instead of 1.5 kA). At this point we would like to stress that using a movable phase mask and a variable polarizing undulator \cite{sasaki:1994} our method allows to independently control OAM and spin, thereby enabling complete control over the total angular momentum carried by an FEL-light beam. 

Compared to other approaches where, e.g. a helical modulator is used to generate OAM beams \cite{hemsing:2011}, our technique is much more general. By modifying the phase mask, the seed laser can be manipulated in order to change the transverse properties of the FEL light at saturation. With a two-quadrant phase mask and a phase jump of $\pi$, a FEL profile close to a TEM$_{01}$ is obtained at high harmonics of the seed laser (not shown). Our technique based on an optical phase mask therefore opens up new opportunities that will allow to manipulate the spatio-temporal properties of modern and future light sources delivering high photon fluxes at XUV and soft x-rays wavelengths.

Finally, we propose to use our scheme in high-order harmonic generation (HHG) in gases. With our setup the generation of well-defined vortex beams with unit topological charge could be extended into the so-called linear propagation regime \cite{garcia:2013}, where the propagation instabilities which lead to formation of low-order vortices \cite{zurch:2012} are not important. Even though the transverse mode selection process is absent in HHG, the propagation effects alone could result in a well-defined dominant $l=1$ OAM mode in far field. Extensive simulations are currently underway to corroborate our hypothesis.

\section{Acknowledgments}
We acknowledge fruitful discussions with Antoine Camper on the general aspects of using a phase mask. We appreciate valuable comments from Simone Di Mitri on the topic of e-beam optics and manipulation. We would also like to thank Sven Reiche and Atoosa Meseck for some clarifications regarding the GENESIS software code.

%

\end{document}